\documentclass{iaus}
\usepackage{graphicx}

\newcommand{\Gyr}{\mbox{Gyr} \ }

\newcommand{\muG}{\mu{\mbox{G}}}

\newcommand{\G}{\mbox{G}}

\begin{document}

\title[Global Cosmic-Ray Driven Dynamo]{Global simulations of galactic dynamo driven by cosmic-rays and exploding magnetized stars}
\author[Hanasz, W\'olta\'nski, Kowalik \& Paw\l{}aszek]{Micha\l \ Hanasz, Dominik W\'olta\'nski, Kacper Kowalik,  
   Rafa\l{} Paw\l{}aszek} 
\affiliation{Centre for Astronomy, Nicholas Copernicus University,
  PL-87148 Piwnice/Toru\'n, Poland, mhanasz@astri.uni.torun.pl}
\pubyear{2009}
\volume{259}  
\pagerange{100--100}
\date{"YOUR MAILING DATE"  and in revised form ??}
\setcounter{page}{119} \jname{Cosmic Magnetic Fields: From Planets,
to Stars and Galaxies} \editors{K.G. Strassmeier, A.G. Kosovichev \&
J.E. Beckman, eds.}
\maketitle
\begin{abstract}
We conduct global galactic--scale magnetohydrodynamical (MHD) simulations of the
cosmic--ray driven dynamo.  We assume that exploding stars  deposit  small--scale,  randomly
oriented, dipolar magnetic fields into the differentially rotating ISM, together
with a portion of cosmic rays, accelerated in supernova shocks.  
Our simulations are performed with the aid of a new parallel MHD code PIERNIK. 
We demonstrate that dipolar magnetic fields supplied on small
SN--remnant scales, can be amplified exponentially by the CR--driven dynamo to the
present equipartition values, and transformed simultaneously to large
galactic--scales by an inverse cascade promoted by resistive processes.
\end{abstract}
%
%
%
%
%
It has been suggested  by Rees (\cite{rees87}) that galactic seed fields 
were created and amplified in stars during early stages of galactic evolution,
and then spread into the interstellar medium (ISM) by stellar explosions, and
subsequently amplified in plerionic (Crab--type) supernova remnants (SNRs). Rees
(\cite{rees87}) estimates that a contribution of $10^6$ randomly oriented
plerionic SNRs may lead to $10^{-9} \G$ mean magnetic fields on galactic
scales.  
%
%
%
The initial setup of our galactic disk is based on the model by
Ferriere~(\cite{ferriere-98}), with the  gravitational potential by
Allen \&  Santill{\'a}n (\cite{allen-santillan-91}). The global CR--driven dynamo model involves basic
elements of local dynamo models presented by Hanasz et al.  
(\cite{hanasz-etal-04}): (1.) Cosmic rays supplied in randomly distributed
SNRs, which are described as relativistic gas diffusing anisotropically along
magnetic field lines, according to the diffusion--advection transport equation,
supplemented to the standard set of resistive MHD equations. (2.) A finite
resistivity of the ISM, responsible for dissipation of small--scale magnetic
fields. Moreover, we assume that no magnetic field is present in the initial
configuration, and that each SN  supplies a weak, randomly oriented, dipolar
magnetic field within the supernova remnant, together with the portion of CRs,
while  the thermal energy output from supernovae is neglected.
%
%
%
Simulations have been performed with the aid of  PIERNIK MHD code (see Hanasz et
al 2009a,b and references therein), which  is a grid--based MPI parallelized, resistive MHD code
based on the Relaxing TVD (RTVD) scheme by Jin \& Xin~(\cite{jin95}) and Pen et
al.~(\cite{pen03}).  The original scheme is extended to deal with the diffusive
CR component  (see Hanasz \& Lesch~(\cite{hanasz-lesch-03})). The simulation has
been performed with the spatial resolution of 1000x1000x160  grid cells, in the
domain spanning 25 kpc x 25 kpc x 8 kpc in $x$, $y$ and $z$ directions,
respectively.
\begin{figure}
\centering
\includegraphics[angle=270,width=0.35\columnwidth]{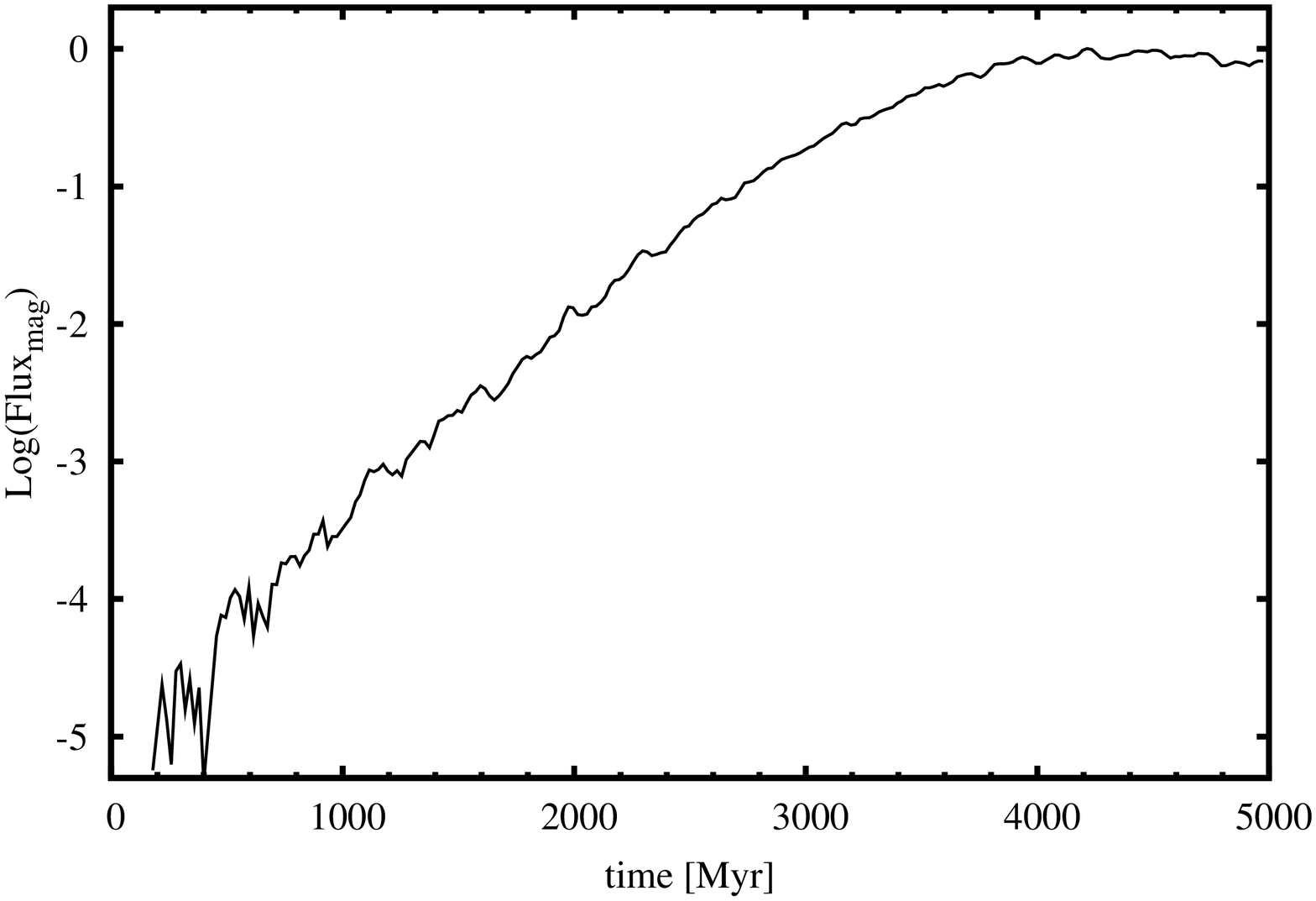} \quad 
\includegraphics[angle=270,width=0.35\columnwidth]{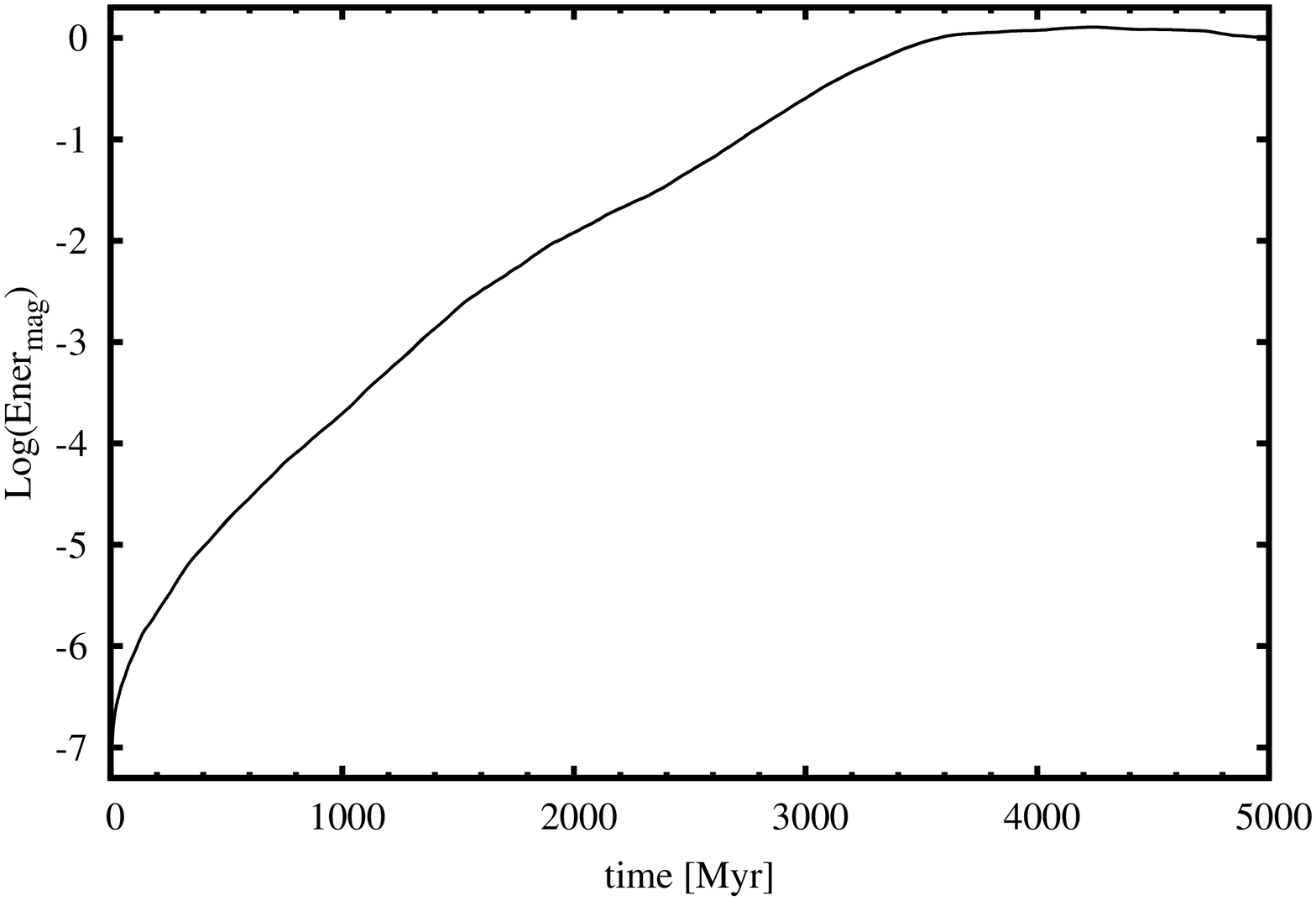}
\caption{Temporal evolution of toroidal magnetic flux (left) and total magnetic energy (right). The final saturation level corresponds to the equipartition magnetic fields}
\label{fig:1}
\end{figure}				
\begin{figure}
  \centerline{\includegraphics[width=0.25\textwidth]{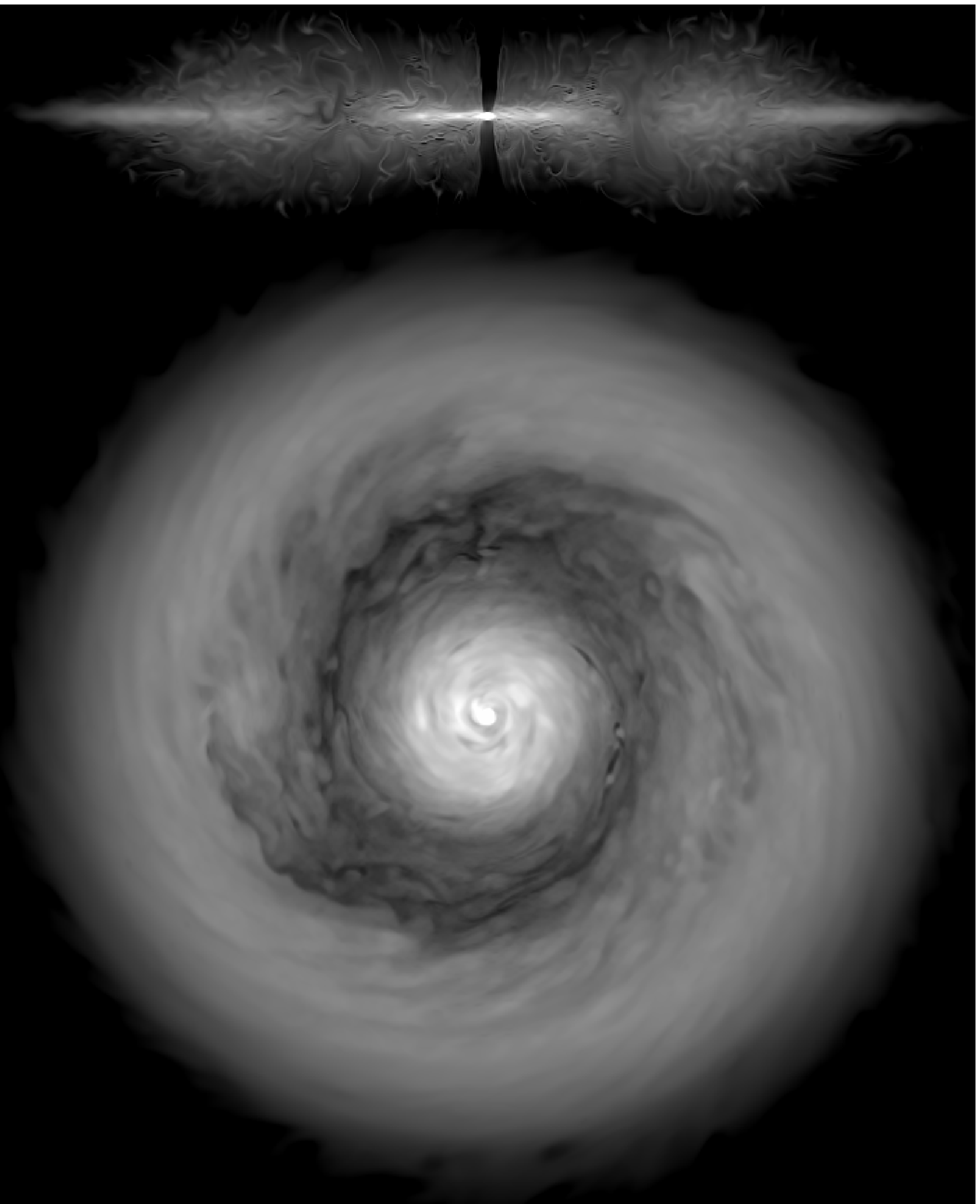} 
	      \includegraphics[width=0.25\textwidth]{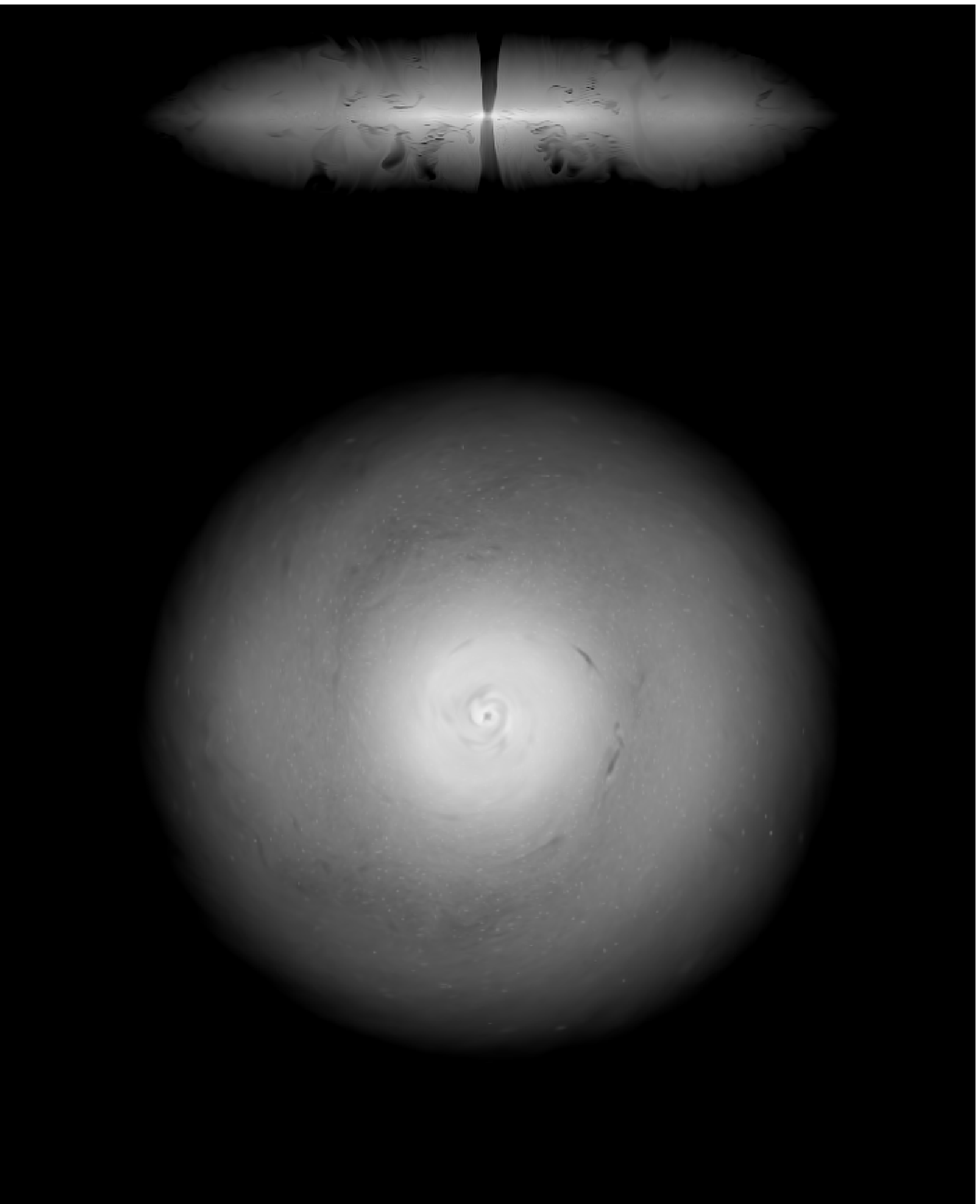}
	      \includegraphics[width=0.25\textwidth]{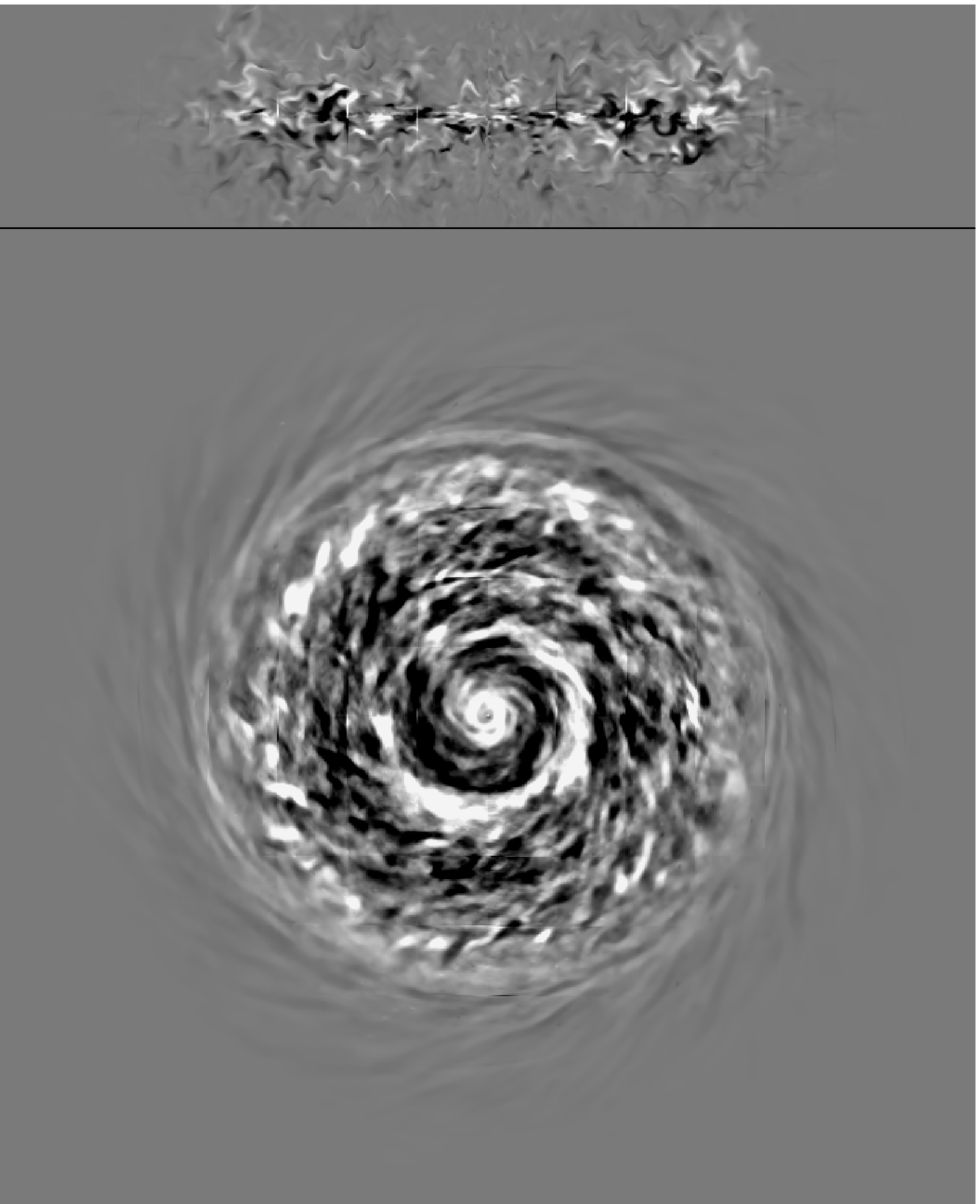} 
	     \includegraphics[width=0.25\textwidth]{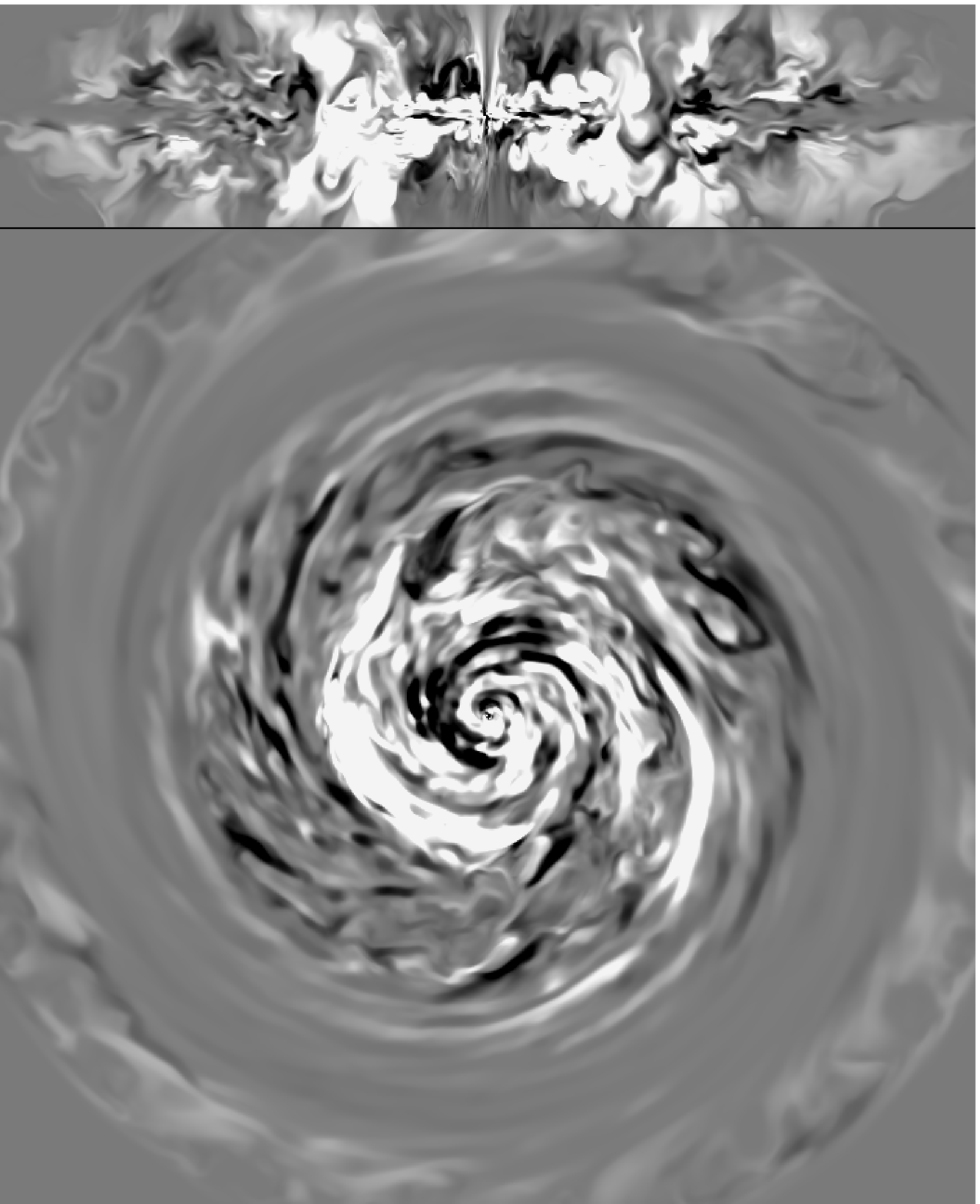}} 
\caption{Logarithm of gas density (1st panel) and cosmic ray energy density 
 (2nd panel) at $t= 4.2 \Gyr$.
 Distribution of toroidal magnetic field at $t=0.5$~Gyr
(3rd panel) and $t= 5$~Gyr (4th panel). Unmagnetized regions of the volume are marked with the gray colour, positive and negative toroidal magnetic fields are marked with white and black colours, respectively. Initially the toroidal magnetic field strength is a few  $10^{-8}\G$ and at the saturated state it is a few $\muG$.}
\label{fig:dens-cren}
\end{figure}				
We show that the CR driven dynamo, seeded by small--scale magnetic dipols and cosmic rays supplied in
supernova remnants, amplifies magnetic fields exponentially by several orders of
magnitude (Fig.~\ref{fig:1}),  up to the saturation level, and develops large scale 
magnetic fields in the disk and the surrounding galactic halo (Fig.~\ref{fig:dens-cren}).  The
horizontal slice demonstrates the spiral structure of the amplified field.
Formation of large lobes of unipolar magnetic fields is apparent in vertical
slices through the disk volume. The magnetic field large--scale structure forms
the X--shaped configuration. The experiment supports strongly the idea by Rees
(1987) that galactic dynamos may have been initiated by small--scale magnetic
fields of stellar origin.
\subsection*{Acknowledgements}
The computations were performed on the GALERA supercomputer in TASK Academic
Computer Centre in Gda\'nsk.   This work was supported by Polish Ministry of
Science and Higher Education through the grant 92/N--ASTROSIM/2008/0.

\end{document}